\documentclass[aps,prl,reprint,superscriptaddress]{revtex4-1}
\usepackage{graphicx}
\usepackage{dcolumn}
\usepackage{bm}
\usepackage{braket}
\usepackage{amsmath}
\usepackage{xcolor}
\usepackage{float}

\definecolor{JPcol}{rgb}{0, 0.4, 0}

\begin{document}

\title{Experimental 11-Dimensional Secret Sharing with Perfect Vortex Beams}

\author{Jonathan Pinnell}
\affiliation{School of Physics, University of the Witwatersrand, Johannesburg 2000, South Africa}
\author{Isaac Nape}
\affiliation{School of Physics, University of the Witwatersrand, Johannesburg 2000, South Africa}
\author{Michael De Oliveira}
\affiliation{School of Physics, University of the Witwatersrand, Johannesburg 2000, South Africa}
\author{Najmeh Tabebordbar}
\affiliation{School of Physics, University of the Witwatersrand, Johannesburg 2000, South Africa}
\author{Andrew Forbes}
\affiliation{School of Physics, University of the Witwatersrand, Johannesburg 2000, South Africa}
\affiliation{Corresponding author: andrew.forbes@wits.ac.za}




\begin{abstract}
Secret sharing is the art of securely sharing information between more than two people in such a way that its reconstruction requires the collaboration of a certain number of parties. Entanglement-based secret sharing schemes which utilise multi-particle entanglement are limited by their scalability. Recently, a high-dimensional single photon secret sharing protocol was proposed which has impressive advantages in scalability. However, the experimental realisation of this protocol remains elusive. Here, by taking advantage of the high-dimensional Hilbert space for orbital angular momentum and using Perfect Vortex beams as their carriers, we present a proof-of-principle implementation of a high-dimensional single photon quantum secret sharing scheme. We experimentally implemented this scheme for 10 participants in $d=$11 dimensions and show how it can be easily scaled to higher dimensions and any number of participants.

\end{abstract}

\maketitle

Quantum cryptography has paved the way in the development of many encryption schemes which utilise features that are unique to quantum mechanics, for example, the no-cloning theorem \cite{wootters1982single}, non-locality \cite{redhead1987incompleteness} and the uncertainty-principle \cite{heisenberg1985anschaulichen} to name a few. These features have fostered the evolution of quantum key distribution (QKD) schemes that are provably secure in the presence of an arbitrarily powerful eavesdropper. In typical QKD schemes, a key is shared between (and is thus restricted to) two parties \cite{bennett2014quantum}. In the past decade, schemes for generating correlated keys shared among multiple parties were developed, namely quantum secret sharing (QSS) protocols \cite{ hillery1999quantum, karlsson1999quantum}.

Traditional QSS schemes were developed with multi-partite entangled quantum states in mind, such as the Greenburg-Horne-Zeilinger (GHZ) state for three parties \cite{hillery1999quantum}. Later, many other schemes emerged including circular QSS \cite{deng2006circular}, dynamic QSS \cite{jia2012dynamic,hsu2013dynamic}, graph state QSS \cite{markham2008graph,bell2014experimental}, verifiable QSS \cite{yang2011verifiable} and QSS based on error correcting QSS \cite{zu2011quantum}. Many of the aforementioned schemes rely on non-local correlations between multiple particles which are difficult to generate and control and cannot yet be transported over appreciable distances.

Interestingly, a new class of QSS schemes involving single photon states has been developed \cite{guo2003quantum} and implemented \cite{schmid2005experimental}. Here, the participants each apply a cascade of local unitaries whilst noting the phases they individually impart on the encoded photon. At the end, based on their choice of imparted phase, the validity of the round is checked after which a subset of the participants can distill the secret. This scheme was initially designed for two dimensional states but was shown to be insecure \cite{he2007comment, qin2008special}. Recently, a high dimensional variation of the single photon QSS protocol was formalised, where the security loop-holes were addressed \cite{tavakoli2015secret}. In this protocol, $d$ mutually unbiased bases (MUBs) \cite{durt2010mutually} are used in the generation and detection of the single photon states, making high dimensional photon encoding and the ability to control each dimension separately of vital importance.

The most common candidate for experimental implementation of single photon QSS schemes is based on the polarization of light \cite{deng2005bidirectional}. However, since polarisation is limited to two dimensions, this implementation restricts scalability. Alternatively, the orbital angular momentum (OAM) degree of freedom of light \cite{yao2011orbital} is an infinite-dimensional Hilbert space and is thus a promising candidate for scalable high dimensional photon encoding processes such as high-dimensional single photon QSS. The OAM of light has also been used to demonstrate the feasibility of high dimensional quantum cryptography \cite{mafu2013highdimQKD}.\\
\begin{figure*}[t]
    \centering
    \includegraphics[width=\textwidth]{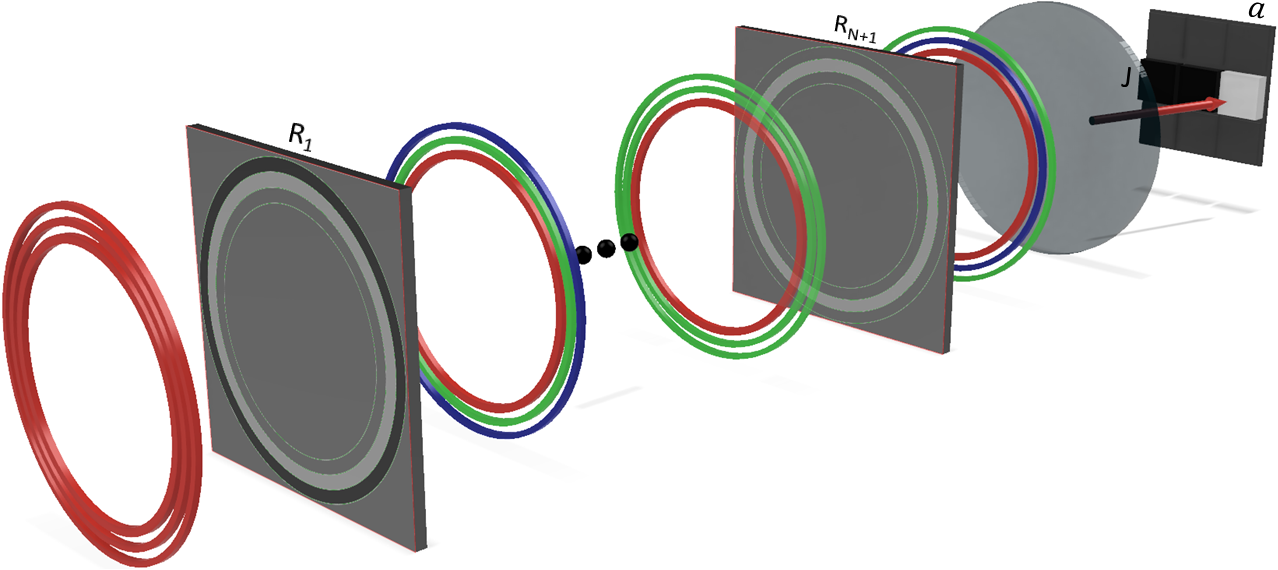}
    \caption{Concept behind our proposed $d$-dimensional secret sharing scheme with Perfect Vortex beams. The distributor generates an attenuated field corresponding to $\ket{e_0^{(0)}}$ which comprises $d$ number of PV rings each containing different OAM values. Each participant applies their unitary in the form of ring apertures encoded on phase-only spatial light modulators (or an equivalent optical device). The final participant transmits the qudit state back to the distributor who deterministically measures in the randomly chosen $J^{th}$ MUB, obtaining outcome $a$. If the round is valid, the distributor's secret can be determined through the collaboration of the remaining participants.}
    \label{fig:concept}
\end{figure*}
Since, high-dimensional single photon secret sharing has only been demonstrated with at most three dimensions using spatial modes of light \cite{smania2016experimental}, here, we outline a proof-of-principle implementation of this scheme with a larger encoding alphabet, utilizing OAM for the basis states and Perfect Vortex (PV) beams as the OAM carriers \cite{vaity2015perfect,Ostrovsky13}. Since a toolbox for deterministic measurement in $d$ dimensions of OAM does not (yet) exist, we utilise a probabilistic OAM measurement based on modal decomposition to project the state. We successfully implement 1 round of this scheme for 10 parties in 11 dimensions. This implementation is easily scalable to higher dimensions and an unlimited number of parties.

The $d$-dimensional secret sharing protocol that we use is based on sequential single qudit communication between $N+1$ participants \cite{tavakoli2015secret}. Each participant operates locally on the single qudit, mapping it to one of the $d$ vectors in one of the $d$ MUBs:
\begin{equation}\label{eq-MUB-Basis}
    \ket{e_{k}^{(j)}}=\frac{1}{\sqrt{d}}\sum_{\ell=0}^{d-1}\omega^{\ell(k+j \ell)}\Ket{\ell} \,,
\end{equation}
where $\ket{e_{k}^{(j)}}$ is the $k^{\text{th}}$ vector in the $j^{\text{th}}$ MUB, $\omega=e^{2\pi i/d}$ and $\ket{\ell}$ represents a vector in the computational basis. The possible values of $\ell$, $j$ and $k$ are integers which lie in the range $[0,d-1]$. It is possible to start with any of the $d^2$ vectors and span the whole $d$-dimensional MUB space by applying the operators 
\begin{align}
    X_d &= \sum^{d-1}_{\ell=0} \omega^\ell\ket{\ell}\bra{\ell} \,, \\
    Y_d &= \sum^{d-1}_{\ell=0} \omega^{\ell^{2}}\ket{\ell}\bra{\ell}\,.
\end{align} 
In words, $X_d$ maps between vectors inside the same MUB while $Y_d$ maps between corresponding vectors in different MUBs. Mathematically, 
\begin{align}
    X_d \ket{e^{(j)}_k} &= \ket{e^{(j)}_{k+1}} \,, \\
    Y_d \ket{e^{(j)}_k} &= \ket{e^{(j+1)}_{k}} \,.
\end{align}
Therefore, by repeatedly and sequentially applying both of these operators, we can span the whole $d$-dimensional MUB space,
\begin{align}
  X_d^x \, Y_d^y \, \ket{e^{j}_k} = \ket{e^{j+y}_{k+x}} \,,
\end{align}
where $x,y \in [0,d-1]$ and the indices are modulo $d$. 

\begin{figure*}[t]
    \centering
    \includegraphics[width=\textwidth]{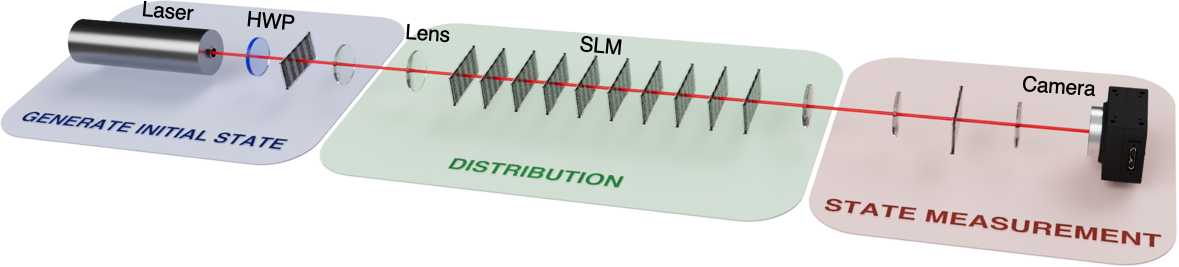}
    \caption{Schematic of the all-digital experimental setup. The $\ket{e_0^{(0)}}$ state is generated from an expanded and collimated He-Ne laser beam using the first SLM. This field is then relayed using a 4f lens system to the first participant who applies their unitary (displayed as a phase map on a SLM) and so on until the last participant whereupon the state is measured through an optical inner product.}
    \label{fig:expSetup}
\end{figure*}

How does the $d$-dimensional secret sharing scheme work? Suppose that there are $N+1$ participants and a distributor $R_{1}$ who desires to share a secret between the $N$ other parties. Firstly, the distributor generates the initial state $\ket{e^{0}_{0}}$ and uniformly samples two random integers $(x_1,y_1)$ from $[0,d-1]$. He/she then locally operates on this qudit state by applying the operator $X^{x_1}_d Y^{y_1}_d$. The distributor then sends this new state $\ket{\psi^d_{R_1}}$ to the next party $R_{2}$. Similarly, this participant also generates two random integers $(x_{2},y_{2})$ from $[0,d-1]$ and locally applies $X^{x_2}_d Y^{y_2}_d$ to the received qudit. The resulting state $\ket{\psi^d_{R_2}}$ is sent to the next participant and so on. Once the qudit has been sequentially communicated between all $N$ parties who each apply $X^{x_n}_d Y^{y_n}_d$, the final participant $R_{N+1}$ then sends the qudit back to $R_{1}$ (the distributor), whereupon the final qudit state is given by,
\begin{align} 
\ket{\psi_{\text{final}}} &= X_d^{x_{N+1}} Y_d^{y_{N+1}} \,\ldots\,X_d^{x_1}Y_d^{y_1} \, \ket{e_0^{(0)}} \, \\
&= \frac{1}{\sqrt{d}} \left( \ket{0} + \sum_{\ell=1}^{d-1} \omega^{\sum_{n=1}^{N+1} (\ell x_n + \ell^2 y_n)} \ket{\ell} \right) \,, \label{eq:final}
\end{align}
Now, $R_{1}$ chooses a random integer $J\in [0,d-1]$ and deterministically measures the qudit in the $J^\text{th}$ MUB with  measurement outcome labelled $a \in \{$0,1,...d-1$ \}$. In random order, parties $R_{2},\ldots,R_{N+1}$ announce their choice of $y_{n}$ whereupon the distributor announces the validity of the round by checking the criterion:
\begin{equation}
    \sum^{N+1}_{n=1}y_n  =  J \,\, \text{mod} \, d \,.
    \label{eq:firstCheck}
\end{equation}
This is effectively verifying whether the distributor measured the qudit in the correct MUB. If the round is valid, the private data $\{x_n\}$ of all the parties satisfy
\begin{equation}
    \sum^{N+1}_{n=1}x_n  =  a \,\, \text{mod} \,  d \,.
    \label{eq:SecCheck}
\end{equation}
If $R_{1}$ changes their private data $x_{1}$ to $x_{1}^{(scrt)}=x_{1}-a $, the set $\{ x_n \}$ exhibit perfect correlations (they sum to 0 modulo $d$) and the $N$ participants can collaborate to determine the secret $x_1^{scrt}$ of the distributor.

In this Letter, we propose an experimental implementation of this protocol based on the OAM states of light. The key steps for an experimental implementation lie in the ability to generate the initial qudit state $\ket{e_0^{(0)}}$, sequentially apply the operators $X_d^{x_n} Y_d^{y_n}$ on the qudit and then deterministically measure the qudit in any of the $d$ MUBs. We'll show, using Perfect Vortex beams as the OAM carriers, how the initial qudit state can be generated, how the unitaries $X_d,Y_d$ can be created and applied and also how the qudit state transforms after each participant. Since a toolbox for deterministic OAM measurement in $d$ dimensions has yet to be developed, we instead utilise a probabilistic mode projection-based measurement. In what follows, we outline the details of this implementation and give experimental results for $d=11$ and $N=9$.

We begin with an overview of Perfect Vortex (PV) beams. Typical vortex modes have a characteristic doughnut shape whose width scales with OAM content; for example, Laguerre-Gaussian modes have a width which scales as $\sqrt{\ell}$ where $\ell \hbar$ is the OAM of a single photon in the field. However, PVs are a set of modes whose field is independent of the OAM that they carry \cite{Ostrovsky13}. These ``special" modes turn out to be the Fourier transform of the well-studied Bessel modes and are described by \cite{vaity2015perfect},
\begin{equation}
PV_R^\ell(r,\phi) \propto \exp \left( - \frac{r^2 + R^2}{T^2} \right) \, I_\ell \left( \frac{2 R r}{T^2} \right) \, \exp(i\ell\phi)\,,
\end{equation} 
where $R,T$ is the radius and thickness of the PV ring and $I_\ell(\cdot)$ is the modified Bessel function.

So, why use PVs for secret sharing? The main aspect is that one can generate a field which is a superposition of PVs that all have different radii,
\begin{equation}
U(r,\phi) \propto \sum_{\ell=0}^{d-1} PV_{R_\ell}^\ell \equiv \sum_{\ell=0}^{d-1} \ket{\ell} \,,
\end{equation}
where $R_\ell$ is the radius of the $\ell^\text{th}$ PV and the thickness of each PV is effectively constant. Since PV fields are independent of OAM, one can then structure the OAM modes in the superposition in a very convenient way (something which cannot be done with any other set of vortex modes). In particular, the PVs in the superposition can be organised so that each ring is spatially separated from the others. It turns out that if the rings are structured such that the radii of adjacent rings $\Delta R$ satisfies $\Delta R \geq 2 T$, then the PVs are sufficiently separated from one another that the OAM modes they carry can be manipulated independently \cite{pinnell2019shaping}. This is key, since it grants one the ability to apply any unitary operation to the qudit OAM state.

From Eq.~\ref{eq:final}, it's apparent that applying the unitary $X_d^{x_n}\,Y_d^{y_n}$ is equivalent to applying specific phase shifts to particular OAM modes in the superposition field. As outlined in \cite{pinnell2019shaping}, these inter-modal phase shifts can be applied in a single step through the use of binary ring apertures encoded on phase-only spatial light modulators (SLMs). Hence, by appropriately choosing the phase within each ring aperture, each participant is able to apply the unitary $X_d^{x_n} Y_d^{y_n}$. 

\begin{figure*}[t]
\centering
\includegraphics[width=\textwidth]{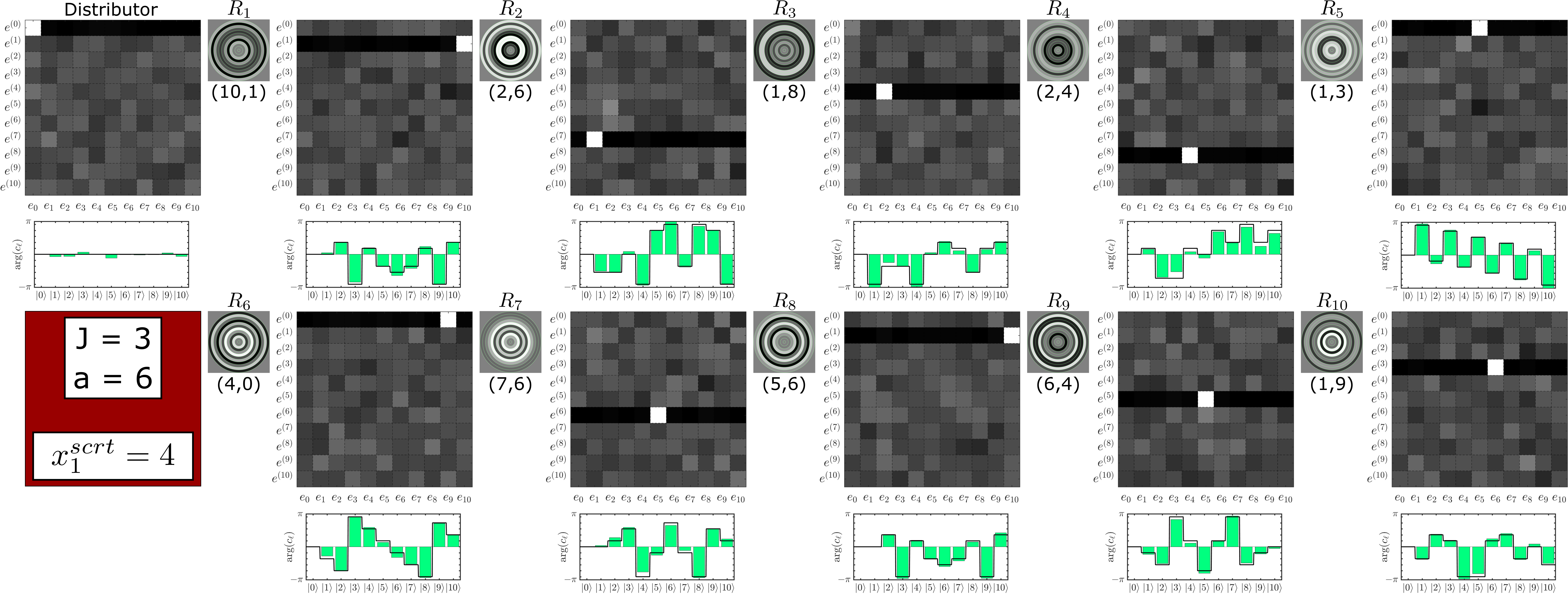}
\caption{Results for 10 participant, $d=11$ dimensional secret sharing with PV beams. Vectors in each MUB are mutually orthogonal which allows one to immediately discern the qudit's state. Vectors in different MUBs have an overlap of $1/\sqrt{d}$. The inter-modal phases are shown below the projection matrices; the dark lines represent the theoretical phases corresponding to the qudit state in $\ket{e_i^{(j)}}$.}
\label{fig:d11}
\end{figure*}

To make the concept concrete, we now outline the steps of this implementation scheme for $d=3$, which is also shown visually in Fig.~\ref{fig:concept}. The initial state/field is given by,
\begin{align}
\ket{e_0^{(0)}} &= \frac{1}{\sqrt{3}} \left( \ket{0} + \ket{1} + \ket{2} \right) \,, \\
&\propto PV_{R_0}^0 + PV_{R_1}^1 + PV_{R_2}^2 \,, \label{eq:PVsuper}
\end{align}
where $R_\ell = R_0 + 3 \ell T$. This superposition means that the OAM modes are ordered sequentially in an increasing manner from the innermost ring and the ring spacing is $\Delta R = 3T$. One can arrange the modes in any desired order, but this choice is the most convenient. The unitary operations $X_3$ and $Y_3$ correspond to,
\begin{align}
X_3 &= \left( \begin{array}{ccc}
1 & 0 & 0 \\
0 & \omega & 0 \\
0 & 0 & \omega^{-1} \end{array} \right)  \,, \\
Y_3 &= \left( \begin{array}{ccc}
1 & 0 & 0 \\
0 & \omega & 0 \\
0 & 0 & \omega \end{array} \right)  \,.
\end{align}
where $\omega = \exp(2\pi i/3)$. For randomly drawn integers $(x_n,y_n)$, the unitary $X_3^{x_n}Y_3^{y_n}$ corresponds to the phase map (modulo $2\pi$),
\begin{equation}
    \Phi = \omega(x_n + y_n) B_{\Delta R}(R_1) + \omega (y_n-x_n) B_{\Delta R}(R_2)\,,
\end{equation}
where $B_{\Delta R}(R_i)$ is the boxcar function centred at $R_i$ and having width $\Delta R$. This phase map is displayed directly onto a SLM to implement the desired unitary, as shown in Fig.~\ref{fig:concept}.

The qudit state is transmitted from participant to participant, who each apply their unitary. Finally, the distributor performs a deterministic measurement in a randomly chosen MUB using some (as yet undeveloped) OAM-MUB mode sorter optic. Once the measurement is performed, shown in Fig.~\ref{fig:concept} as a projection matrix where each element is a state in the $d^2$ MUB state space, the classical post processing steps then follow as usual.

The PV field, states, MUBs and unitaries extend analogously for $d=11$. A schematic of the experimental setup used to implement the scheme for $10$ participants is shown in Fig.~\ref{fig:expSetup}. The first SLM was used to generate the field corresponding to the $\ket{e_0^{(0)}}$ state from an expanded and collimated He-Ne laser beam. The unitaries (phase maps) of each participant were applied in sequence, whereupon the final state was measured using an optical inner product. To ensure consistency, we use two different methods of measuring the state: a projection matrix approach and a modal decomposition approach. We resorted to probabilistic/statistical measurements since there does not yet exist a toolbox for deterministic measurement of OAM-MUBs for arbitrary dimension $d$. Each element of the projection matrices corresponds to performing the optical overlap between the MUB vector $\ket{e_i^{(j)}}$ and the shared qudit state. The modal decomposition corresponds to the overlap between the qudit state and the two superposition states $\ket{0} + \ket{\ell}$ and $\ket{0} + i\ket{\ell}$. Performing these two optical overlaps is known to be sufficient for determining the inter-modal phase between $\ket{0}$ and $\ket{\ell}$ \cite{Forbes2016} and is now also known to be effective for PV beams \cite{Pinnell:19}. The projection matrix route of reconstructing the state requires $d^2$ measurements, whilst the modal decomposition route requires $2d-1$ measurements. The experimental results for 1 round of QSS are summarised in Fig.~\ref{fig:d11}; we performed a state measurement after each participant to show the evolution of the qudit state. We can read off the final state from the final projection matrix, whereupon we see that the round is valid provided $J=3$. In this case, the distributor's measurement outcome is $a = 6$ and so the shared secret is $x_1 - a = 4$.

Extending the scheme to any dimension is straightforward: add more rings. In principle, the limitation when utilising SLMs is the number of PV rings that can fit onto the screen. Making the rings thinner would allow one to pack more rings onto the SLM. However, it turns out that the optical system's numerical aperture limits how thin the PV rings can be made. Since PVs are the Fourier transform of Bessel beams, these fields are not propagation invariant. This means that PVs have to be relayed from plane to plane using an imaging system. The thickness of the PV is inversely related to the width of the Bessel beam since $T = 2f/k w_0$ where $w_0$ is the Gaussian width of the Bessel-Gaussian beam. Hence, making the rings thinner causes the corresponding Bessel beam to be larger and there will come a point where the optical system will not be able to collect all the necessary light.

In summary, we proposed a scalable implementation of a high dimensional quantum secret sharing protocol which utilises MUB states of single photons carrying OAM. We confirmed the efficacy of our scheme in a proof-of-principle experiment in 11-dimensions with 10 participants using probabilistic state measurement. Since the protocol relies on a deterministic measurement of the photon's qudit state, future work should investigate the development of a toolbox to achieve this with MUBs of OAM.

\bibliography{Ref.bib}


\end{document}